\documentclass[12pt]{article}

\usepackage{a4wide}

\usepackage{amssymb}
\usepackage{latexsym}

\usepackage{epsfig}
\usepackage{graphicx}

\usepackage{amsmath,amsthm,amsfonts,amssymb}

\newcommand{\be}{\begin{equation}}
\newcommand{\ee}{\end{equation}}
\newcommand{\bea}{\begin{eqnarray}}
\newcommand{\eea}{\end{eqnarray}}
\newcommand{\beqar}{\begin{eqnarray*}}
\newcommand{\eeqar}{\end{eqnarray*}}
\newcommand{\nn}{\nonumber}

\catcode`\@=11

\@addtoreset{equation}{section}

\def\@normalsize{\@setsize\normalsize{15pt}\xiipt\@xiipt
\abovedisplayskip 14pt plus3pt minus3pt%
\belowdisplayskip \abovedisplayskip
\abovedisplayshortskip  \z@ plus3pt%
\belowdisplayshortskip  7pt plus3.5pt minus0pt}
\def\small{\@setsize\small{13.6pt}\xipt\@xipt
\abovedisplayskip 13pt plus3pt minus3pt%
\belowdisplayskip \abovedisplayskip
\abovedisplayshortskip  \z@ plus3pt%
\belowdisplayshortskip  7pt plus3.5pt minus0pt
\def\@listi{\parsep 4.5pt plus 2pt minus 1pt
            \itemsep \parsep
            \topsep 9pt plus 3pt minus 3pt}}

\def\underline#1{\relax\ifmmode\@@underline#1\else
        $\@@underline{\hbox{#1}}$\relax\fi}
\@twosidetrue \relax

\catcode`@=12

\catcode`\@=11

\def\section{\@startsection{section}{1}{\z@}{3.5ex plus 1ex minus
   .2ex}{2.3ex plus .2ex}{\large\bf}}

\def\ps@headings{\def\@oddfoot{}\def\@evenfoot{}
\def\@oddhead{\hbox{}\hfill
        \makebox[.5\textwidth]{\raggedright\ignorespaces --\thepage{}--
        \hfill }}
\def\@evenhead{\@oddhead}
\def\subsectionmark##1{\markboth{##1}{}}
}

\ps@headings

\catcode`\@=12

\begin{document}
%%%%%%%%%%%%%%%%%%%%%%%%%%%%%%%%%%%%%%%%%%%%%%
%\begin{figure}[h]
%\centering
%\includegraphics[scale=0.5]{a:figure1p.ps}
%\end{figure}
%%%%%%%%%%%%%%%%%%%%%%%%%%%%%%%%%%%%%%%%%%%%%%

\begin{titlepage}

\rightline{hep-th/0502169} \rightline{February 2005}

\begin{centering}
\vspace{2cm}
{\large {\bf The Need of Dark Energy for Dynamical Compactification of Extra Dimensions on the Brane}}\\

\vspace{1.5cm}

 {\bf Bertha~Cuadros-Melgar}$^{a,*}$ and {\bf Eleftherios~Papantonopoulos}$^{b,**}$ \\
\vspace{.2in}

$^{a,b}$ National Technical University of Athens, Physics
Department, \\ Zografou Campus, GR 157 80, Athens, Greece. \\
\vspace{3mm} $^{a}$ Instituto de Fisica, Universidade de S\~{a}o
Paulo  C. P. 66.318, \\ CEP 05315-970, S\~{a}o Paulo, Brazil.

\end{centering}
\vspace{3cm}

\begin{abstract}

We consider a six-dimensional braneworld model and we study the
cosmological evolution of a (4+1) brane-universe. Introducing
matter on the brane we show that the scale factor of the physical
three-dimensional brane-universe is related to the scale factor of
the fourth dimension on the brane, and the suppression of the
extra dimension compared to the three dimensions requires the
presence of dark energy.

\end{abstract}

\vspace{3cm}
\begin{flushleft}

$^{*}~$ e-mail address: bertha@fma.if.usp.br  \\
$ ^{**}$ e-mail address: lpapa@central.ntua.gr

\end{flushleft}
\end{titlepage}

\section{Introduction}

Recent high quality data from various independent observations
like the cosmic microwave background
anisotropies~\cite{Spergel:2003cb}, large scale galaxy
surveys~\cite{Scranton:2003in} and type IA
 supernovae~\cite{Riess:1999ka,Goldhaber:2001ux} suggest that
 most of the energy content  of our universe is in the form of dark matter and dark
energy. Although there have been many plausible explanations for
these dark components, it is challenging
 to try to explain these exotic ingredients of the universe using
 alternative gravity theories.

 Braneworld models are higher-dimensional modified gravity theories, which share many common
features with general relativity, but
 at the same time give corrections to the conventional gravity theory such as modifications to the
 Newton's law and alternative non-conventional cosmology. The essence of the braneworld scenario
 is that matter and gauge interactions are  localized on a
three-dimensional hypersurface (called brane) in a
higher-dimensional spacetime while  gravity propagates in all
spacetime
 (called bulk). This idea gained momentum
 the last years~\cite{AADD,randall} because of its connection with string theory
  (for a review see~\cite{Maartens:2003tw}).  The
 cosmology of these and other related models
with one transverse to the brane extra dimension is well
 understood (for a review see \cite{reviews}).
In the cosmological generalization of \cite{randall},
 the early times (high energy limit) cosmological evolution is modified
by the square of the matter density on the brane, while the bulk
leaves its imprints on the brane by the ``dark radiation" term
 \cite{bdl,RScosmology,binetruy,csaki,maartens}. The presence of a bulk cosmological constant
 in \cite{randall} gives conventional
cosmology at late times (low energy limit)
\cite{bdl,RScosmology,binetruy}.

In braneworld scenarios, contrary to the Kaluza-Klein theories,
the extra dimensions can be large if the geometry is non trivial.
If the hierarchy problem is addressed in the braneworld scenarios
for example, the extra dimensions should be large~\cite{AADD}. In
a cosmological context however, these extra dimensions are
observationally much smaller than the size of our perceived
universe. Initially the universe could have started with the sizes
of all dimensions at the Planck length. Then, a successful
cosmological model should accommodate in a natural way a mechanism
by which the extra dimensions remained comparatively small during
cosmological evolution.

Such a mechanism was proposed in~\cite{Brandenberger:1988aj}. The
basic idea is that strings dominate the dynamics of the early
universe and can see each other most efficiently in 2(p+1) (p=1
for strings) dimensions. Therefore, strings can only interact in 3 spatial dimensions, while strings moving in higher dimensions
eventually cease to interact efficiently and their winding modes
will prevent them from further expanding. If branes are included,
it was shown in~\cite{Alexander:2000xv} that strings will still
dominate the evolution of the universe at late times so the
mechanism of~\cite{Brandenberger:1988aj} still survives.

In this work we show that in a (4+1) braneworld model in a
six-dimensional spacetime bulk with a cosmological constant, the
unknown form of brane dark energy is responsible for the dynamical
compactification of the extra fourth dimension relatively to the
physical three dimensions, and during the various cosmological
evolution phases of the brane-universe it keeps the extra
dimension frozen. This model can be generalized to higher
dimensions at the expense of calculational difficulties. Six or
higher-dimensional braneworld models were considered as
generalizations of the Randall-Sundrum model.
In~\cite{Kanti:2001vb} static and non-static solutions in a
6-dimensional bulk as well as the stability of the radion field
were discussed, while in~\cite{Arkani-Hamed:1999gq} the evolution
of the extra dimensions transverse to the 3-brane in a Kasner-like
metric was studied.

In this paper we follow a different approach. First we consider a
4-brane fixed at some position in the sixth dimension and we
derive the dynamical six-dimensional bulk equations in normal
gaussian coordinates with a cosmological constant in the bulk and
considering matter on the brane~\cite{binetruy}. We look for time
dependent solutions allowing for two scale factors, the usual
scale factor $a(t)$ of the three dimensional space and a scale
factor $b(t)$ for the extra fourth dimension. If $a(t)=b(t)$ we
get the Friedmann equation of the generalized Randall-Sundrum
model in six dimensions describing a four dimensional
universe~\cite{Abdalla:2002ir}. If $a(t) \neq b(t)$ we get a
generalized Friedmann equation in six dimensions~\cite{BCM-PhD}.

The problem can be looked at a different angle. If $a=b$, we can
write the bulk six-dimensional metric in ``Schwarzschild"
coordinates and then we have the equivalent description of a
4-brane moving in a static Schwarzschild-(A)dS six-dimensional
bulk~\cite{csaki,Cuadros-Melgar:2003zh}. If however $a \neq b$, a
brane observer uses $a$ and $b$, for whom they are static
quantities, to measure the departure from six-dimensional
spherical symmetry of the bulk. The important result of this
consideration is that, since the brane observer needs to define a
cosmic time in order to derive an effective Friedmann equation on
the brane~\cite{csaki}, $a$ and $b$ are related through the
Darmois-Israel junctions conditions and because of that, their
relation depends on the energy-matter content of the brane. The
physical reason of the existence of such a relation between $a$
and $b$ is that the requirement of having a cosmological evolution on
the brane introduces a kind of compactification on it and
the relation between $a$ and $b$ acts as a constraint of the brane
motion in the bulk.

Using the six-dimensional generalized Friedmann equation and
assuming that $p \neq \hat{p} $, where $p$ is the pressure of the
physical three dimensions and $\hat{p}$ corresponds to the fourth dimension,
we make a systematic numerical study of the cosmological evolution
of the scale factors $a(t)$ and $b(t)$ for several values of the
parameters of the model, $\Lambda_{6}$ the
six-dimensional cosmological constant, $k$ the brane spatial
curvature and $w$ and $ \hat{w}$ parameterizing the form of the
brane energy-matter content of the three dimensions and of the
extra fourth dimension respectively. We find that, in order the
fourth dimension to be small relatively to the other three
dimensions and to remain constant during cosmological evolution, $
\hat{w}$ must be negative, indicating the presence of a dark form
of energy in the extra fourth dimension. We find this result for
all cases considered, (A)dS or Minkowski bulk, open, closed or
flat universe, radiation, dust, cosmological constant and dark
energy dominated universe and the specific value of $ \hat{w}$
depends on the energy-matter content of the other three
dimensions.

The paper is organized as follows. After the introduction in Sec.
1, in Sec. 2 we derive the generalized Friedmann equation of a
4-brane in a six-dimensional bulk. We followed two approaches, that of
a static brane in a dynamical bulk and that of a moving brane in a static
bulk. The second approach gives us information of how the two
scale factors are related. In Sec. 3 we make a numerical
investigation of how the two scale factors evolve under
various choices of the parameters of the model, and in Sec. 4 our
results are summarized.

\section{A Cosmological Braneworld Model in
Six Dimensions}\label{intro6d}

In this section we describe a braneworld cosmological model in six
dimensions. This model is a generalization of the existing
braneworld models in five-dimensions. We study the model following
two different approaches. First we use normal gaussian coordinates
to describe a static 4-brane at fixed position in a
six-dimensional spacetime bulk. The six-dimensional Einstein
equations are derived and with the use of the appropriate junction
conditions the generalized Friedmann equation is given. The
cosmological evolution described by this generalized Friedmann
equation involves the usual three-dimensional scale factor $a(t)$
and the scale factor $b(t)$ describing the cosmological evolution
of the extra fourth dimension.

We consider next a dynamical brane moving in a bulk described by
six-dimensional static ``coordinates". In this case the dynamical
brane is moving on a geodesic which is given by the junctions
conditions. We derive the equations of motion of the brane which
for a brane observer describe the cosmological evolution on the
brane. We discuss the connection between the static and dynamical
brane models and the physical information that can be extracted
from these approaches.

\subsection{Static Brane in a Dynamical Bulk} \label{6ddynamic}

 We look for a solution to the Einstein equations in six-dimensional spacetime with a metric of the form
\be \label{metric6} ds^2 = -n^2(t,y,z) dt^2 + a^2(t,y,z)
d\Sigma_{k} ^2 + b^2(t,y,z)  dy^2 + d^2(t,y,z) dz^2 \, , \ee where
$d\Sigma_{k} ^2$ represents the 3-dimensional spatial sections
metric with $k=-1,\,0,\,1$ corresponding to the hyperbolic, flat
and elliptic spaces, respectively.

If the brane is fixed at the position $z_{0}$, then the total
energy-momentum tensor can be decomposed in two parts
corresponding to the bulk and the brane as
\begin{equation}\label{emtensor}
\tilde T^M _N = \breve T^{M(B)} _N + T^{M (b)} _N \, ,
\end{equation}
where the energy-momentum tensor on the brane is
\begin{equation}\label{brane}
T^{M (b)} _N = { {\delta (z-z_0)} \over {d}} \,diag \,(-\rho,
p,p,p,\hat p, 0) \, ,
\end{equation} where $\hat p$ is the pressure in the extra brane
dimension. We assume that there is no matter in the bulk and the
energy-momentum tensor of the bulk is proportional to the
six-dimensional cosmological constant.

The non-zero Einstein's tensor components are given by
\bea\label{g00}
 G_{00}&=&
3 { \dot a \over a} \left( {\dot a \over a} +
    {\dot b \over
    b} + {\dot d \over d} \right) + {{\dot b \dot d} \over {bd}}
+ {n^2 \over b^2}
    \left\{ {{\partial_y b \partial_y d}\over {bd}} - 3 {{\partial_y a \partial_y d}\over {ad}} + 3 {{\partial_y a \partial_y b}\over {ab}} - \right.\nonumber\\&&\left.-3 {{(\partial_y a)^2}\over a^2} - 3 {{\partial_y ^2 a} \over a} - {{\partial_y ^2 d} \over d} \right\}
+ {n^2 \over d^2} \left\{ 3 {{\partial_z a \partial_z d}\over {ad}} - 3 {{\partial_z a \partial_z b}\over {ab}} + {{\partial_z b \partial_z d}\over {bd}} - \right.\nonumber\\
&&\left.- 3 {{(\partial_z a)^2}\over {a^2}} - 3 {{\partial_z ^2
a}\over {a}} - {{\partial_z ^2 b}\over {b}} \right\} + 6k {n^2
\over a^2} \eea

%%%%%%%%%%%%%%%%%%%%%%%%%%%%%%%%%%%%%%%%%%%%%%%%%%%%%%%%%%%%%%%%%%%%%

\bea\label{gij}
G_{ij}&=& \left\{ {a^2 \over n^2} \left( - {{\dot a^2} \over a^2} -2 {\ddot a \over a} - {\ddot b \over b} - {\ddot d \over d} + 2 {{\dot a \dot n}\over{an}} - 2 {{\dot a \dot b}\over{ab}} - 2 {{\dot a \dot d}\over{ad}} + {{\dot b \dot n}\over{bn}} + {{\dot d \dot n}\over{dn}} - {{\dot b \dot d}\over{bd}} \right) + \right. \nonumber \\
&&+{a^2 \over b^2} \left( -2 {{\partial_y a \partial_y b}\over{ab}} + 2 {{\partial_y a \partial_y n}\over{an}} + 2 {{\partial_y a \partial_y d}\over{ad}} - {{\partial_y b \partial_y n}\over{bn}} + {{\partial_y d \partial_y n}\over{dn}} - \right. \nonumber \\
&&\left. - {{\partial_y b \partial_y d}\over{bd}} + {{(\partial_y a)^2}\over a^2} + 2 {{\partial_y ^2 a}\over a} +{{\partial_y ^2 d}\over d} + {{\partial_y ^2 n}\over n}  \right) + {a^2 \over d^2} \left( -2 {{\partial_z a \partial_z d}\over{ad}} + \right.\nonumber \\
&&\left. + 2 {{\partial_z a \partial_z b}\over{ab}} + 2 {{\partial_z a \partial_z n}\over{an}} + {{\partial_z b \partial_z n}\over{bn}} - {{\partial_z d \partial_z n}\over{dn}} - {{\partial_z b \partial_z d}\over{bd}} + {{(\partial_z a)^2}\over a^2} + \right.\nonumber \\
&&\left. \left. +2 {{\partial_z ^2 a}\over a} + {{\partial_z ^2
n}\over n} + {{\partial_z ^2 b}\over b}  \right)
\right\}\gamma_{ij} - 3 k \gamma_{ij} \eea

%%%%%%%%%%%%%%%%%%%%%%%%%%%%%%%%%%%%%%%%%%%%%%%%%%%%%%%%%%%%%%%%%%%%%%%%%

\bea\label{g55}
G_{55}&=&  {b^2 \over n^2} \left( - 3 {{\dot a^2} \over a^2} - {\ddot d \over d} - 3 {\ddot a \over a} -3 {{\dot a \dot d}\over{ad}} + {{\dot d \dot n}\over{dn}} + {{\dot a \dot n}\over{an}} \right) + {{\partial_y d \partial_y n}\over{dn}} + 3 {{\partial_y a \partial_y n}\over{an}} +  \nonumber \\
&& +3 {{\partial_y a \partial_y d}\over{ad}} + 3 {{(\partial_y a)^2}\over a^2} + {b^2 \over d^2} \left( - {{\partial_z d \partial_z n}\over{dn}} + 3 {{\partial_z a \partial_z n}\over{an}} - 3 {{\partial_z a \partial_z d}\over{ad}} + \right. \nonumber \\
&&\left.  + 3 {{(\partial_z a)^2}\over a^2} + 3  {{\partial_z ^2
a}\over a} + {{\partial_z ^2 n}\over n} \right)  - 3k {b^2 \over
a^2}\eea

%%%%%%%%%%%%%%%%%%%%%%%%%%%%%%%%%%%%%%%%%%%%%%%%%%%%%%%%%%%%%%%%%%%%%%%%%%%

\bea\label{g66}
G_{66}&=& {{d^2}\over n^2} \left( -3 {{\dot a^2}\over {a^2}} - 3{\ddot a \over a} - {\ddot b \over b} - 3 {{\dot a \dot b}\over {ab}} + 3 {{\dot a \dot n}\over {an}} +  {{\dot b \dot n}\over {bn}} \right) + {{d^2}\over b^2} \left( -3 {{\partial_y a \partial_y b}\over{ab}} -  \right. \nonumber \\
&& \left. -{{\partial_y b \partial_y n}\over{bn}} + 3 {{\partial_y a \partial_y n}\over{an}} + 3 {{(\partial_y a)^2}\over a^2} + 3 {{\partial_y ^2 a}\over a} + {{\partial_y ^2 n}\over n} \right) + 3 {{\partial_z a \partial_z b}\over{ab}} + \nonumber \\
&&+ 3 {{\partial_z a \partial_z n}\over{an}} + {{\partial_z b
\partial_z n}\over{bn}} +3 {{(\partial_z a)^2}\over a^2} - 6 k
{d^2 \over a^2} \eea

%%%%%%%%%%%%%%%%%%%%%%%%%%%%%%%%%%%%%%%%%%%%%%%%%%%%%%%%%%%%%%%%%%%%%%%%%%%

\bea\label{g05} G_{05}&=& -3 {{\partial_y \dot a}\over {a}} -
{{\partial_y \dot d}\over {d}} + 3 {{\dot a \partial_y n}\over
{an}} + 3 {{\dot b \partial_y a}\over {ba}} +  {{\dot d \partial_y
n}\over {dn}} +  {{\dot b \partial_y d}\over {bd}} \eea

%%%%%%%%%%%%%%%%%%%%%%%%%%%%%%%%%%%%%%%%%%%%%%%%%%%%%%%%%%%%%%%%%%%%%%%%%%%

\bea\label{g06}
 G_{06}&=& -3 {{\partial_z \dot a}\over {a}} - {{\partial_z \dot b}\over {b}} + 3 {{\dot a \partial_z n}\over {an}} + 3 {{\dot d \partial_z a}\over {da}} +  {{\dot b \partial_z n}\over {bn}} +  {{\dot d \partial_z b}\over {db}}
\eea

%%%%%%%%%%%%%%%%%%%%%%%%%%%%%%%%%%%%%%%%%%%%%%%%%%%%%%%%%%%%%%%%%%%%%%%%%%%

\bea\label{g56} G_{56}&=& - {{\partial_z \partial_y n}\over n} - 3
{{\partial_z \partial_y a}\over a} + {{\partial_z b \partial_y
n}\over{bn}} + {{\partial_z n \partial_y
  d}\over{nd}} + 3 {{\partial_z b \partial_y a}\over{ba}} + 3 {{\partial_z a \partial_y d}\over{ad}}
\eea

%%%%%%%%%%%%%%%%%%%%%%%%%%%%%%%%%%%%%%%%%%%%%%%%%%%%%%%%%%%%%%%%%%%%%%%%%%%

where $\gamma_{ij}$ is the maximally symmetric three-dimensional
metric.

The presence of the brane in $z_{0}$ imposes boundary conditions
on the metric: it must be continuous through the brane, while its
derivatives with respect to $z$ can be discontinuous at the brane
position. This means that the generated Dirac delta function in
the metric second derivatives with respect to $z$ must be matched
with the energy-momentum tensor components (\ref{brane}) to
satisfy the Einstein equations~\cite{bdl}. Therefore, using
(\ref{g00}), (\ref{gij}) and (\ref{g55}) we obtain the following
Darmois-Israel conditions,
\begin{eqnarray}\label{israel}
{{[\partial_z a]} \over {a_0 d_0}} &=& -{{\kappa_{(6)} ^2} \over
4}
(p-\hat p + \rho) \, ,\nonumber \\
{{[\partial_z b]} \over {b_0 d_0}} &=& -{{\kappa_{(6)} ^2} \over
4}
\left\{ \rho - 3(p-\hat p) \right\} \, ,\\
{{[\partial_z n]} \over {d_0 n_0}} &=& {{\kappa_{(6)} ^2} \over 4}
\left\{ \hat p + 3 (p+ \rho) \right\} \, , \nonumber
\end{eqnarray}
where the subscript $(0)$ indicates quantities on the brane. The
energy conservation equation on the brane can be derived taking
the jump of the $(06)$ component of the Einstein equations, using
Eq.~(\ref{g06})
\begin{eqnarray}\label{dist1}
[G_{06}] &=& -{3 \over {a_0}} [\partial_z \dot a] - {1 \over b_0}
[\partial_z \dot b] + 3 {{\dot a_0} \over {a_0 n_0}} [\partial_z
n] + {{\dot d_0}\over {d_0}} \left\{ 3 {{[\partial_z a]} \over
a_0} + {{[\partial_z b]}\over b_0} \right\} + {{\dot b_0}\over
{b_0}} {{[\partial_z n]} \over n_0}
\end{eqnarray}
Using the junction conditions (\ref{israel}) and the corresponding
time derivatives given by
\begin{eqnarray}
\partial_t \left( {{[\partial_z a]}\over {a_0 d_0}} \right) &=&
-{{\kappa_{(6)} ^2}\over 4} (\dot p -\dot{\hat p} + \dot\rho) \, ,\\
\partial_t \left( {{[\partial_z b]}\over {b_0 d_0}} \right) &=&
-{{\kappa_{(6)} ^2}\over 4} \left\{\dot\rho -3(\dot p -\dot{\hat
p}) \right\} \, ,
\end{eqnarray}
we obtain
\begin{equation}\label{conserva6}
\dot\rho + 3 (p+\rho) {{\dot a_0}\over a_0} + (\hat p + \rho)
{{\dot b_0}\over b_0} =0 \, .
\end{equation}

To find the Friedmann equation we take the jump of the $(66)$
component of the Einstein equations (Eq.~(\ref{g66})),
\begin{equation}
3 {{[(\partial_z a)^2]} \over {a_0 ^2}} + 3 {{[\partial_z a
\partial_z b]}\over {a_0 b_0}} + 3 {{[\partial_z a \partial_z
n]}\over {a_0 n_0}} + {{[\partial_z b \partial_z n]}\over {b_0
n_0}} = 0 \, .
\end{equation}
We use the fact that
\begin{equation}\label{prodsal}
[\partial_z f \, \partial_z g] = \#\partial_z f\# \, \,
[\partial_z g] + [\partial_z f] \,\, \#\partial_z g\# \, ,
\end{equation}
where $$\#f(y)\# = {{f(0^+) + f(0^-)}\over 2}\, ,$$ is the mean
value of the function $f$ through $y=0$, and we arrive to the
following equation
\begin{equation}\label{valmed}
{{\# \partial_z a \#}\over a_0} p = {1 \over 3} \rho {{\#
\partial_z n \#}\over n_0} - {1 \over 3} \hat p {{\# \partial_z b
\#}\over b_0} \, .
\end{equation}

We take now the mean value of the Eq.~(\ref{g66})
\begin{eqnarray}
{{d_0 ^2}\over{n_0 ^2}} \left( - 3 {{\dot a_0 ^2}\over {a_0 ^2}}
-3 {{\ddot a_0}\over a_0} -
 {{\ddot b_0}\over b_0} - 3
 {{\dot a_0 \dot b_0}\over{a_0 b_0}} + 3 {{\dot a_0 \dot n_0}\over{a_0
 n_0}} + {{\dot b_0 \dot n_0}\over{b_0 n_0}}  \right) + {{d_0 ^2}\over{b_0 ^2}} \left(  -3 {{\partial_y a_0 \partial_y b_0}\over{a_0 b_0}} - \right. && \nonumber \\
\left. -{{\partial_y b_0 \partial_y n_0}\over{b_0 n_0}} + 3 {{\partial_y a_0 \partial_y n_0}\over{a_0 n_0}} + 3 {{(\partial_y a_0)^2}\over {a_0 ^2}} + 3 {{\partial_y ^2 a_0}\over a_0} + {{\partial_y ^2 n_0}\over n_0} \right) = && \nonumber \\
=-3 {{\# (\partial_z a) ^2\#} \over {a_0 ^2}} -3 {{\#\partial_z a
 \,\,\partial_z b \#}\over {a_0 b_0}} - 3 {{\#\partial_z a
 \,\,\partial_z n \#}\over {a_0 n_0}} - {{\#\partial_z b
 \,\,\partial_z n \#}\over {b_0 n_0}} + &&\nonumber \\
+6k{{d_0 ^2}\over {a_0 ^2}}+\kappa_{(6)} ^2 \breve T_{66} \, .&&
\end{eqnarray}
Analogously to the equation (\ref{prodsal}), we have
\begin{equation}\label{prodvalmed}
\#\partial_z f \,\,\partial_z g\# = \#\partial_z f\# \#\partial_z
g\# + {1\over 4} [\partial_z f] [\partial_z g] \, .
\end{equation}
Thus, using (\ref{prodvalmed}), the junction equations,
(\ref{valmed}) and the fact that $\#\partial_z a\#=\#\partial_z
b\#=0$ (we have assumed a $Z_2$ symmetry), we arrive to the
generalized Friedmann equation in six-dimensions
\begin{eqnarray}\label{friedmanndist}
\left( {{\ddot a_0}\over a_0} + {1\over 3} {{\ddot b_0}\over b_0}
+ {{\dot a_0 \dot b_0}\over{a_0 b_0}} + {{\dot a_0 ^2}\over {a_0
^2}} \right) &=&
 - {{\kappa_{(6)} ^4}\over {32}} \left\{ \rho (\rho +2p +
{2\over 3} \hat p) + (p-\hat p)^2 \right\} -\nonumber \\&&-
2{k\over {a_0 ^2}}
  -{{\kappa_{(6)} ^2}\over{3 d_0 ^2} } \breve
T_{66}\, ,
\end{eqnarray} where we have assume $(\partial_{y}a)_{0}=0$, $(\partial_{y}^{2}a)_{0}=0$
and we have chosen $n_0=1$.

In the case of a 3-brane in a five-dimensional bulk, the first
integral of the space-space component of the Einstein equations,
with the help of the other equations can be done
analytically~\cite{binetruy} and this results in the Friedmann
equation on the brane with the dark radiation term as an
integration constant~\cite{maartens}. In our case the Einstein
equations (\ref{g00})-(\ref{g56}) cannot be integrated
analytically and therefore, the usual form of the Friedmann
equation on the brane cannot be extracted from
(\ref{friedmanndist}). Nevertheless, if $a(t)$ and $b(t)$ are
related, this equation will give the cosmological evolution of the
scale factor $a(t)$.

If $a(t)=b(t)=\cal{R}$(t) then (\ref{friedmanndist}) becomes
\begin{equation} \label{rsgen}
2 {{\ddot {\cal R}}\over {\cal R}} + 3 \left( {{\dot {\cal
R}}\over
    {\cal R}} \right)^2 = -3 {{\kappa_{(6)} ^4}\over {64}} \rho ^2 -
    {{\kappa_{(6)} ^4}\over {8}} \rho p - 3 {{k}\over {{\cal R}^2}} - {{\kappa_{(6)} ^2}\over {2 d_0 ^2}} \breve T^6 _6 \, .
    \end{equation}
This equation is the generalization of the Randall-Sundrum
Friedmann-like equation in six dimensions, and it has, as
expected, the $\rho^{2}$ term with a coefficient adjusted to six
dimensions. Note that this equation can easily be generalized to D
dimensions.

\subsection{Dynamical Brane in a Static Bulk}\label{6dstatic}

We consider a 4-brane moving in a six-dimensional
Schwarzschild-AdS spacetime. The metric in the ``Schwarzschild"
coordinates can be written as
\begin{equation}\label{bhmetric}
ds^2 = - h(z) dt^2 + {z^2 \over l^2} d\Sigma_k ^2 + h^{-1} (z)
dz^2 \, ,
\end{equation}
where
\begin{equation}\label{spacediff}
d\Sigma_k ^2 = {{dr^2} \over {1-kr^2}} + r^2 d\Omega_{(2)} ^2 +
(1-kr^2) dy^2 \, ,
\end{equation}
and
\begin{equation}
h(z) = k + {z^2 \over l^2} - {M \over z^3} \,  .\label{hsch}
\\\end{equation}
Comparing the metric (\ref{metric6}) with the metric
(\ref{bhmetric}) we can make the following identifications
\begin{eqnarray}\label{identifica}
n(z) &=& \sqrt{h(z)} \, , \nonumber \\
a(z)&=&b(z)=z/l~, \nn \\ d(z)&=& \sqrt{h^{-1}(z)} \, .
\end{eqnarray}
The two approaches of a static brane in a dynamical bulk described
in Sec.~\ref{6ddynamic} and of a moving 4-brane in a static bulk
are equivalent~\cite{Mukohyama:1999wi}. To prove this equivalence,
 consider the Darmois-Israel conditions for a moving brane,
$[K^\mu _\nu] = -\kappa_{(6)} ^2 \left(T^\mu _\nu - {1 \over 4} T
h^\mu _\nu \right)$, with $h^\mu _\nu$ being the induced metric on
the brane; analogously to the case in 5 dimensions~\cite{csaki} one can obtain~\cite{BCM-PhD}
\begin{eqnarray}
{{-h'-2 \ddot{\cal R}}\over {\sqrt{h+ \dot{\cal R}^2}}} &=&
-\kappa_{(6)}
^2 ({3\over 4} \rho -p) \, , \label{equ1} \\
-2 {{\sqrt{h+\dot{\cal R}^2}} \over {\cal R}} &=&
{{\kappa_{(6)}^2}
  \over 4} \rho \, ,\label{equ2}
\end{eqnarray} where we have defined $z=\mathcal{R}(t)$.
Combining these two equations and using (\ref{hsch}) we find that
\begin{equation} \label{brfriedeq}
2 {{\ddot {\cal R}}\over {\cal R}} + 3 \left( {{\dot {\cal
R}}\over
    {\cal R}} \right)^2 = -3 {{\kappa_{(6)} ^4}\over {64}} \rho ^2 -
    {{\kappa_{(6)} ^4}\over {8}} \rho p - 3 {{k}\over {{\cal R}^2}} -
    {5\over {l^2}} \, .
\end{equation}
Comparing with (\ref{rsgen}) we see that
\begin{eqnarray}
{\kappa_{(6)} ^2} \breve T^6 _6 ={{\kappa_{(6)} ^2}\over {d ^2}}
\breve T_{66} = {{10}\over l^2} \nonumber \\
\Rightarrow l^{-2} = -{{\kappa_{(6)} ^2}\over {10}} \Lambda_{_{6}}
\, ,
\end{eqnarray}
with $\breve T^6 _6 =-\Lambda_{6}$ the bulk cosmological constant,
and $l$ the size of the AdS space. Therefore, a brane observer
describes the cosmological evolution of a four-dimensional
universe with the Friedmann equation (\ref{brfriedeq}) with
$\mathcal{R}$ parameterizing the motion of the brane in the $z$
direction.

To describe the motion of the 4-brane with $a(z)\neq b(z)$ we can
write the metric (\ref{bhmetric}) as
 \begin{equation}\label{bhmetric1}
ds^2 = - n(z)^{2} dt^2 +a^{2}(z)d\Sigma_3 ^2 +b(z)^{2}dy^2+
d^{2}(z)dz^2\, .
\end{equation} We
denote the position of the brane
  at any bulk time $t$ by $z = {\cal R}(t)$ as before. Then, an observer on the brane defines the proper time
from the relation \be n^2(t,{\cal R}(t))\dot{t}^{2}-d^2(t,{\cal
R}(t)) \dot{\cal R}^2=1~, \label{dtdtau} \ee which ensures that
the induced metric on the brane will be in FRW form
\begin{eqnarray}\label{induced}
ds^2_{induced} &=& - \left[ n^2(t,{\cal R}(t)) \dot
t^2-d^2(t,{\cal R}(t)) \dot
{\cal R}^2 \right] d\tau ^2 \nonumber \\ &&+ \,\,a^2(t,{\cal R}(t)) d\Sigma^2 _{(3)}+b^{2}(t,{\cal R}(t))dy^2 \nonumber \\
&=& -d\tau^2 + a^2(t,{\cal R}(t)) d\Sigma^2 _{(3)}+b^{2}(t,{\cal
R}(t))dy^2 \, ,
\end{eqnarray}
where the dot indicates derivative with respect to the brane time
$\tau$. The extrinsic curvature tensor on the brane is given by
\begin{equation}\label{excurv6}
K_{MN} = h^L _M \bigtriangledown _L n_N \, ,
\end{equation}
where $nΒ$ is a unitary vector field normal to the brane and
\begin{equation}\label{eta}
h_{MN} = g_{MN} - n_M n_N~,
\end{equation}
is the induced metric on the brane. To calculate the components of
$nΒ$, we use the relations
\begin{equation}\label{rel}
g_{MN} n^M n^N =1 \quad , \quad g_{MN}  n^M u^N =0 \, ,
\end{equation}
where we introduced the unitary velocity vector corresponding to
the brane,
\begin{equation}\label{u}
u^A = \left\{ {{dt}\over {d\tau}},
0,0,0,0,{{d\cal{R}}\over{d\tau}} \right\} \, .
\end{equation} Then we find
\begin{equation}\label{ntilde}
n^A = \Big{(} -nd \dot{\cal R},0,0,0,0,nd\dot{t}\Big{)} \,.
\end{equation}
The components of the extrinsic curvature (\ref{excurv6}) using
the induced metric (\ref{induced}) and (\ref{ntilde}) read
\begin{eqnarray}
K_{\tau\tau}&=&-nd\dot{t}\ddot{\cal R}+nd\dot{\cal R}\ddot{t}
-{{\dot t \partial_z n}\over {d}} + \dot{\cal R}^2 \dot t \left( d \partial_z n - n \partial_z d \right) \, , \label{ttcomp} \\
K_{ij}&=&\frac{a\partial_{z}a}{d}\sqrt{1+d^{2}\dot{\cal
R}^{2}}\gamma_{ij}~, \label{ijcomp} \\
K_{55}&=&\frac{b\partial_{z}b}{d}\sqrt{1+d^{2}\dot{\cal R}^{2}}~.
\label{55comp}
\end{eqnarray}
Eliminating $\ddot{t}$ and $\dot{t}$ from (\ref{ttcomp}), using
(\ref{dtdtau}), equation (\ref{ttcomp}) becomes \bea
K_{\tau\tau}=\frac{d^{2} \dot{d} \dot{\cal R}^{3}-d\ddot{\cal R}}
{\sqrt{1+d^{2}\dot{\cal R}^{2}}}-\frac{\sqrt{1+d^{2}\dot{\cal
R}^{2}}}{n}\Big{(}d\dot{n}\dot{\cal
R}+\frac{\partial_{z}n}{d}-(d\partial_{z}n-n\partial_{z}d)\dot{\cal{R}}^{2}
\Big{)}.
\end{eqnarray}
Introducing an energy-momentum tensor on the brane
\begin{equation}
\hat{T}_{\mu\nu}=h_{\nu\alpha}T^{\alpha}_{\mu}-\frac{1}{4}T
h_{\mu\nu}, \end{equation} where
$T^{\alpha}_{\mu}=diag(-\rho,p,p,p,\hat{p})$, the Darmois-Israel
conditions
\begin{equation}
[K_{\mu\nu}]=-\kappa^{2}_{(6)}\hat{T}_{\mu\nu}
\end{equation}
give the equations of motion of the brane
\begin{eqnarray}
\frac{d^{2} \dot{d} \dot{\cal R}^{3}-d\ddot{\cal R}}
{\sqrt{1+d^{2}\dot{\cal R}^{2}}}&-&\frac{\sqrt{1+d^{2}\dot{\cal
R}^{2}}}{n}\Big{(}d\dot{n}\dot{\cal
R}+\frac{\partial_{z}n}{d}-(d\partial_{z}n-n\partial_{z}d)\dot{\cal{R}}^{2}
\Big{)}\nonumber \\ &=&
-\frac{\kappa^{2}_{(6)}}{8}\Big{(}3(\rho+p)+\hat{p} \Big{)}~,\label{dyneqs} \\
\frac{\partial_{z}a}{ad}\sqrt{1+d^{2}\dot{\cal R}^{2}}&=&
-\frac{\kappa^{2}_{(6)}}{8}\Big{(}\rho+p-\hat{p} \Big{)}~, \label{const1} \\
 \frac{\partial_{z}b}{bd}\sqrt{1+d^{2}\dot{\cal R}^{2}}&=&
-\frac{\kappa^{2}_{(6)}}{8}\Big{(}\rho-3(p-\hat{p})\label{const2}
\Big{)}~.
\end{eqnarray}
Notice that if $a=b={\cal R}/l$, we recover equations (\ref{equ1})
and (\ref{equ2}), corresponding to the Schwarzschild-AdS
spacetime.

Equation (\ref{dyneqs}) is the main dynamical equation describing
the movement of the brane-universe in the six-dimensional bulk,
while a combination of (\ref{const1}) and (\ref{const2}) acts as a
constraint relating $a$ and $b$ (remember that for a brane
observer $a$ and $b$ are static, depending only on z)\be
a=\mathcal{A}~b^{(\rho+p-\hat{p})/(\rho-3(p-\hat{p}))}\label{constrel},\ee
where $\mathcal{A}$ is an integration constant. This is the main
result of our paper: because of the Darmois-Israel conditions, the
relative cosmological evolution of $a$, the scale factor of the
three-dimensional physical universe and $b$, the scale factor of
the extra dimension, depends on the dynamics of the energy-matter
content of the brane-universe. In the next section we analyze the
cosmological evolution of the brane-universe assuming various
forms of energy-matter on the brane.

Another interesting observation is that, a brane observer can
measure the departure from full six-dimensional spherical symmetry
of the bulk using the quantities $a$ and $b$. If $a(z)=b(z)=z/l$
then the symmetry of the bulk is $S^{4}$ having a six-dimensional
Schwarzschild-(A)dS black hole solution. If $a(z)\neq b(z)$ fixing
$a$ to be $a(z)=z/l$, because of (\ref{constrel}) $b(z)$ is given
by \be b=\left\{ \frac{1}{\mathcal{A}}
\Big{(}\frac{z}{l}\Big{)}\right\}^{(\rho-3(p-\hat{p}))/(\rho+p-\hat{p})}
\label{inva} \ee and we expect the topology of the bulk to be
$S^{3}\times {\cal M}$,  ${\cal M}$ being a compact or non-compact
manifold. There are no analytical solutions in the six-dimensional
spacetime with such topology and the reflection of this on the
brane is the difficulty of the brane equations
(\ref{dyneqs})-(\ref{const2}) to be integrated analytically. Note
also that because of (\ref{inva}), a change in the topology of the
bulk is triggered by the dynamics of the matter distribution on
the brane.

\section{The Cosmological Evolution of the Brane-Universe}

To study the cosmological evolution of the four-dimensional
brane-universe, we made the following assumptions for the initial
conditions and the matter distribution on the brane. We assume
that the universe started as a four-dimensional one at the Planck
scale, all the dimensions were of the Planck length and the matter
was isotropically distributed. In this case the cosmological
evolution is described by the generalized Friedmann equation
(\ref{rsgen}). Then an anisotropy was developed in the sense that
$\hat{p}=Qp$ with $Q\neq 1$. The cosmological evolution is now
described by (\ref{friedmanndist}) supplemented with the
constrained equation (\ref{constrel}) and the matter distribution
on the brane is given
by the equations of state \bea p&=&w \rho,\label{eqstq} \\
\hat{p}&=& \hat{w} \rho. \label{eqstqhat} \eea Using
(\ref{constrel}) and relations (\ref{eqstq}) and (\ref{eqstqhat})
the generalized Friedmann equation (\ref{friedmanndist}) becomes
\bea
\frac{\hat{w}-w}{1+\hat{w}}~\dot{H}_{a}&+&\frac{(1+\hat{w})(-3w+2\hat{w}-1)+3(1+\hat{w})^{2}}{(1+\hat{w})^{2}}
~H^{2}_{a} \nn \\
&+&\frac{2w-\hat{w}+1}{(1+\hat{w})^{2}}\frac{\dot{\rho}}{\rho}~H_{a}+\frac{2+\hat{w}}{3(1+\hat{w})^{2}}
\frac{\dot{\rho}^{2}}{\rho^{2}}-\frac{\ddot{\rho}}{3(1+\hat{w})\rho}
\nn \\ &=&
 - {{\kappa_{(6)} ^2}\over {32}} \left\{ 1+2w+\frac{2}{3}\hat{w}+(w-\hat{w})^{2} \right\}
 \rho^{2} -2{k\over {a^2}}
  +{{\kappa_{(6)} ^2}\over{3 } }
\Lambda_{6}\, . \eea Using the conservation equation
(\ref{conserva6}) to eliminate $\rho$ and its derivatives from the
above equation, the cosmological evolution of the
three-dimensional scale factor $a(t)$ is given by the equation
\bea \Big{[}1+\frac{B}{3}\Big{]}\ddot{a}a^{2C+1}&+&\Big{[}
\frac{B^{2}}{3}+\frac{2B}{3}+1 \Big{]}\dot{a}^{2}a^{2C}+
\frac{\kappa_{(6)} ^2} {32} a^{2} \Big{[}
 1+2w+\frac{2}{3}\hat{w}+(w-\hat{w})^{2} \Big{]} \nn \\
 &-& a^{2C} \Big{[} a^{2} \frac{\kappa_{(6)}^2}{3} \Lambda_{6}-2k
 \Big{]}=0~, \label{aevolut} \eea  where the constants $B$ and $C$ are given by
  \bea B&=&\frac{1-3w+3\hat{w}}{1+w-\hat{w}}~, \\
C&=&3(1+w)+B(\hat{w}+1)~, \eea while the $b(t)$ scale factor is
\be
 b(t)=a(t)^{B}~. \label{bevolut}
\ee We will make a numerical analysis of equations (\ref{aevolut})
and (\ref{bevolut}) and study the time evolution of the two scale
factors for different backgrounds and spatial brane-curvature. We
will allow for all possible forms of energy-matter on the physical
three dimensions ($w$=0, 1/3, -1/3) and also for the possibility
of dark energy ($w$=-1), leaving $\hat{w}$ as a free parameter. We
note here that the aim of this work is not to present a detailed
six-dimensional braneworld model, but rather to see if in
braneworld there is in operation a mechanism for the suppression
of the extra dimensions compared to the three physical dimensions
under different forms of energy-matter on the brane.

For the scale factor $b(t)$ to be small compared to the scale
factor $a(t)$, the constant $B$ in (\ref{bevolut}) should be
negative. In Table 1 we give the allowed range of values of
$\hat{w}$ for various values of $w$. These values in turn were
used to plot the time evolution of the scale factors $a(t)$ and
$b(t)$ using (\ref{aevolut}) and (\ref{bevolut}) respectively. The
criterion for the acceptance of a solution is to give a growing
evolution of $a(t)$ and a decaying and freezing out evolution for
$b(t)$. The results for various choices of the parameters of the
model are presented in the following.

%%%%%%%%%%%%%%%%%%%%%%%%%%%%%%%%%%%%%%%%%%%%%5
\begin{table}[h]
\begin{center}
\begin{tabular}{|l|l|r|}
 \hline
$~~~w$ & $~~~~~~~~~~\hat{w}$ \\
\hline
$\,\,-1$  & $>0\,\,\ ~~~or \,\,\,\ <-4/3 $ \\
\hline
$\,\,\,\,~0$ & $>1\,\,\ ~~~or \,\,\ <-1/3 $\\
\hline
$\,\,\,1/3$ & $>4/3\,\,\ or \,\,\,\ <0 $\\
\hline
$-1/3$ & $>2/3\,\,\ or \,\,\,\ <-2/3 $\\
\hline
\end{tabular}
\end{center}
\caption{The allowed values of $\hat{w}$ for $B$ to be negative.}
\end{table}
%%%%%%%%%%%%%%%%%%%%%%%%%%%%%%%%%%%%%%%%%%%%%

%%%%%%%%%%%%%%%%%%%%%%%%%%%%%%%%%%%%%%%%%%%%%
\begin{figure}[h]
\centering
\includegraphics[scale=0.6,angle=-90]{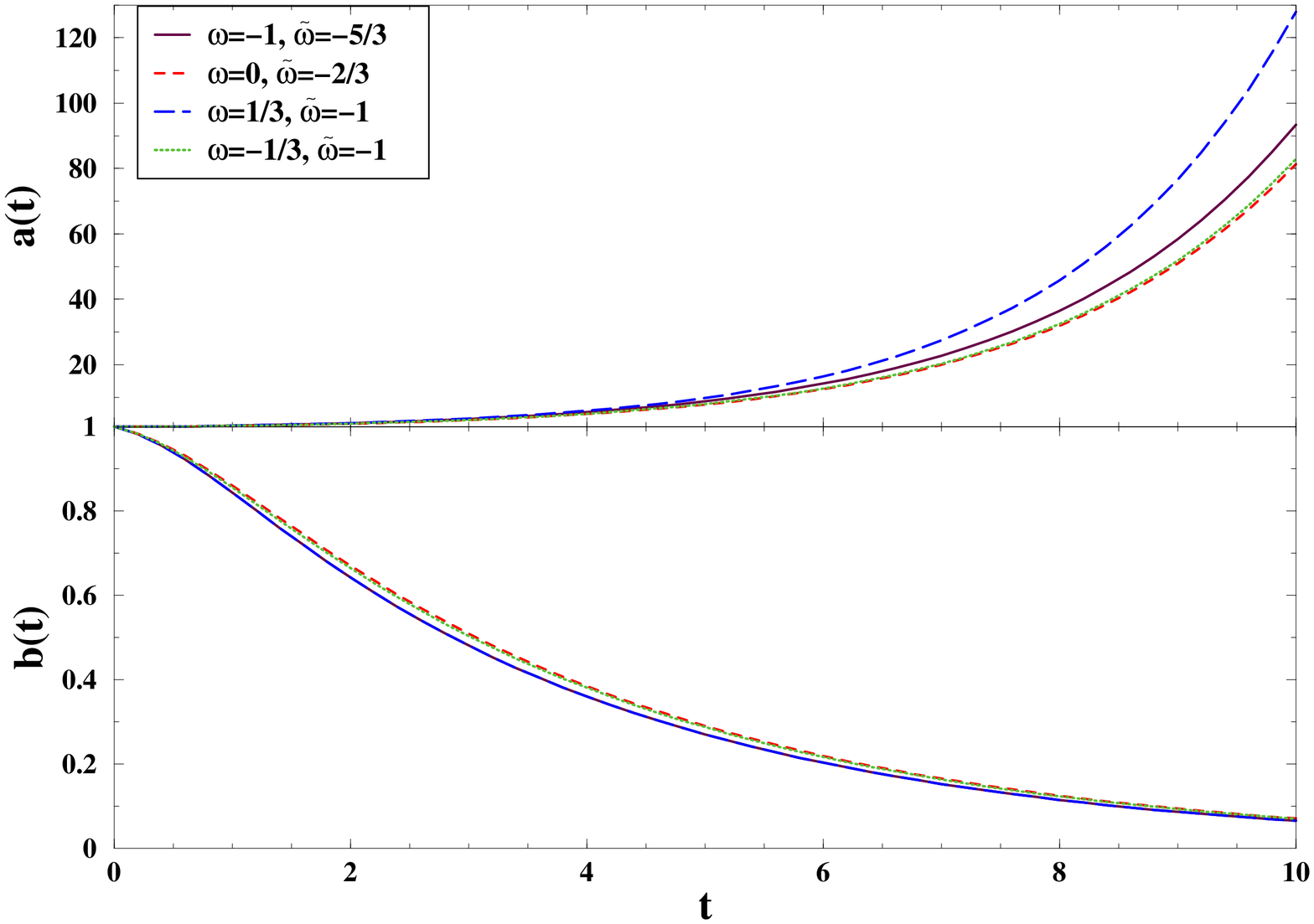}
\caption{Time evolution of the scale factors $a(t)$ and $b(t)$ for
$\Lambda_{6}>0$ and $k=0$.}
\end{figure}
%%%%%%%%%%%%%%%%%%%%%%%%%%%%%%%%%%%%%%%%%%%%%%

%%%%%%%%%%%%%%%%%%%%%%%%%%%%%%%%%%%%%%%%%%%%%
\begin{figure}[h]
\centering
\includegraphics[scale=0.6,angle=-90]{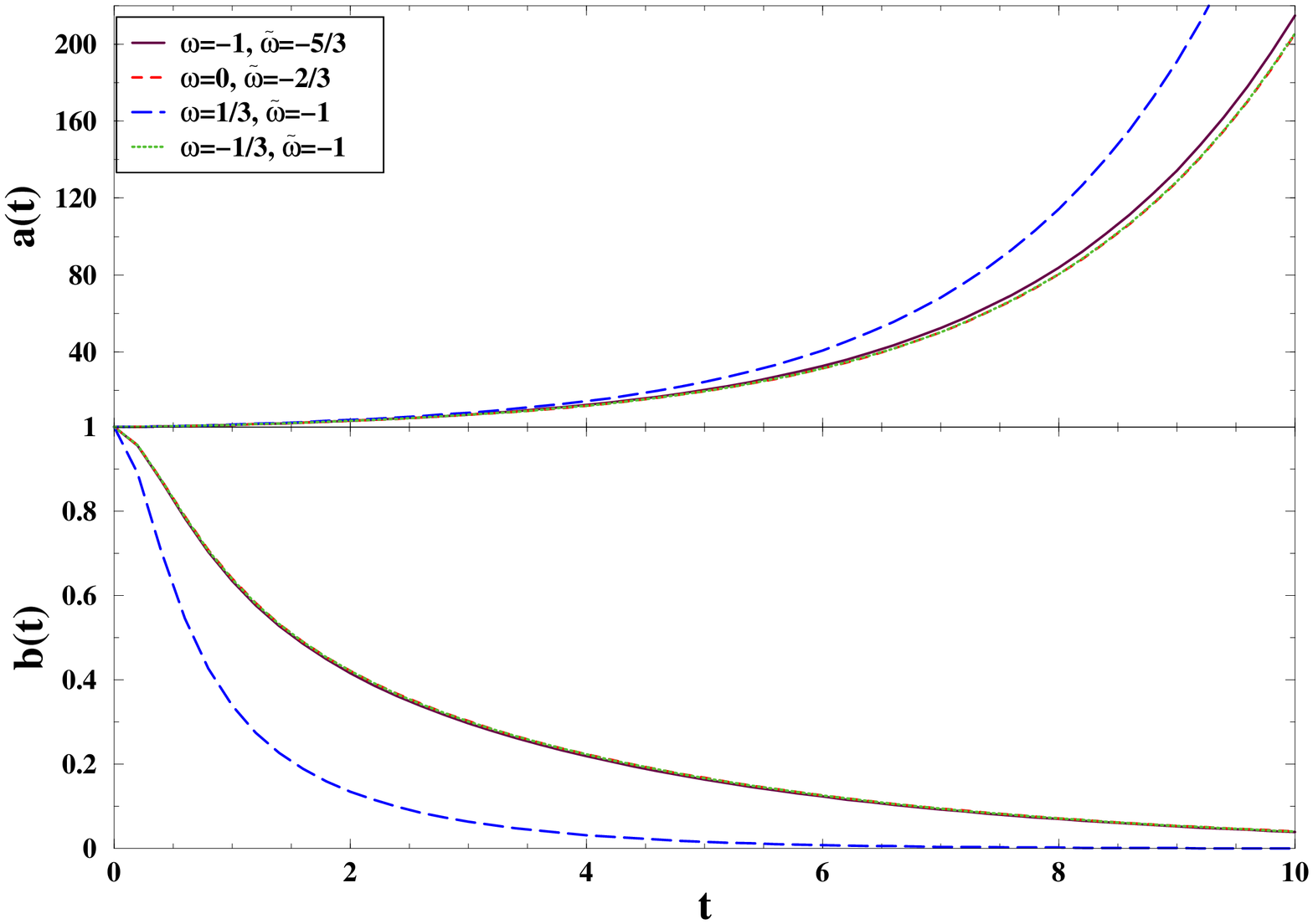}
\caption{Time evolution of the scale factors $a(t)$ and $b(t)$ for
$\Lambda_{6}>0$ and $k=-1$.}
\end{figure}
%%%%%%%%%%%%%%%%%%%%%%%%%%%%%%%%%%%%%%%%%%%%%%

\subsection{De Sitter Bulk}

For $\Lambda_{6}>0$ we find that only negative values of $\hat{w}$
according to Table 1 give acceptable solutions for
$k=0,-1$, while the $k=1$ solutions are not acceptable for any value of
$\hat{w}$. A typical evolution of the two scale factors is shown
in Fig. 1 and Fig. 2, for various values of $w$.

In both cases the scale factor $a(t)$ grows very fast, while the
scale factor of the fourth dimension goes very fast to small
values where it stays constant for the whole evolution. The reason
for the fast growth of $a(t)$ is that the cosmological constant of
the bulk $\Lambda_{6}$ acts as an effective cosmological constant
on the brane and drives an exponential growth. This is common to
the braneworld models with a cosmological constant in the bulk. It
also happens in the five-dimensional Randall-Sundrum model if
we do not impose the fine tuning between the brane tension and the
five-dimensional cosmological constant.

\subsection{Anti De Sitter Bulk}

For $\Lambda_{6}<0$ and $k=0,1$ we do not get any acceptable
solution. The scale factors either go to infinity, or $a(t)$ grows
and $b(t)$ decays for a while and at some point they interchange
behaviours, crossing each other and going to infinity. However,
for $k=-1$ there is an interesting evolution of an oscillating
universe shown in Fig. 3.

%%%%%%%%%%%%%%%%%%%%%%%%%%%%%%%%%%%%%%%%%%%%%
\begin{figure}[h]
\centering
\includegraphics[scale=0.6,angle=-90]{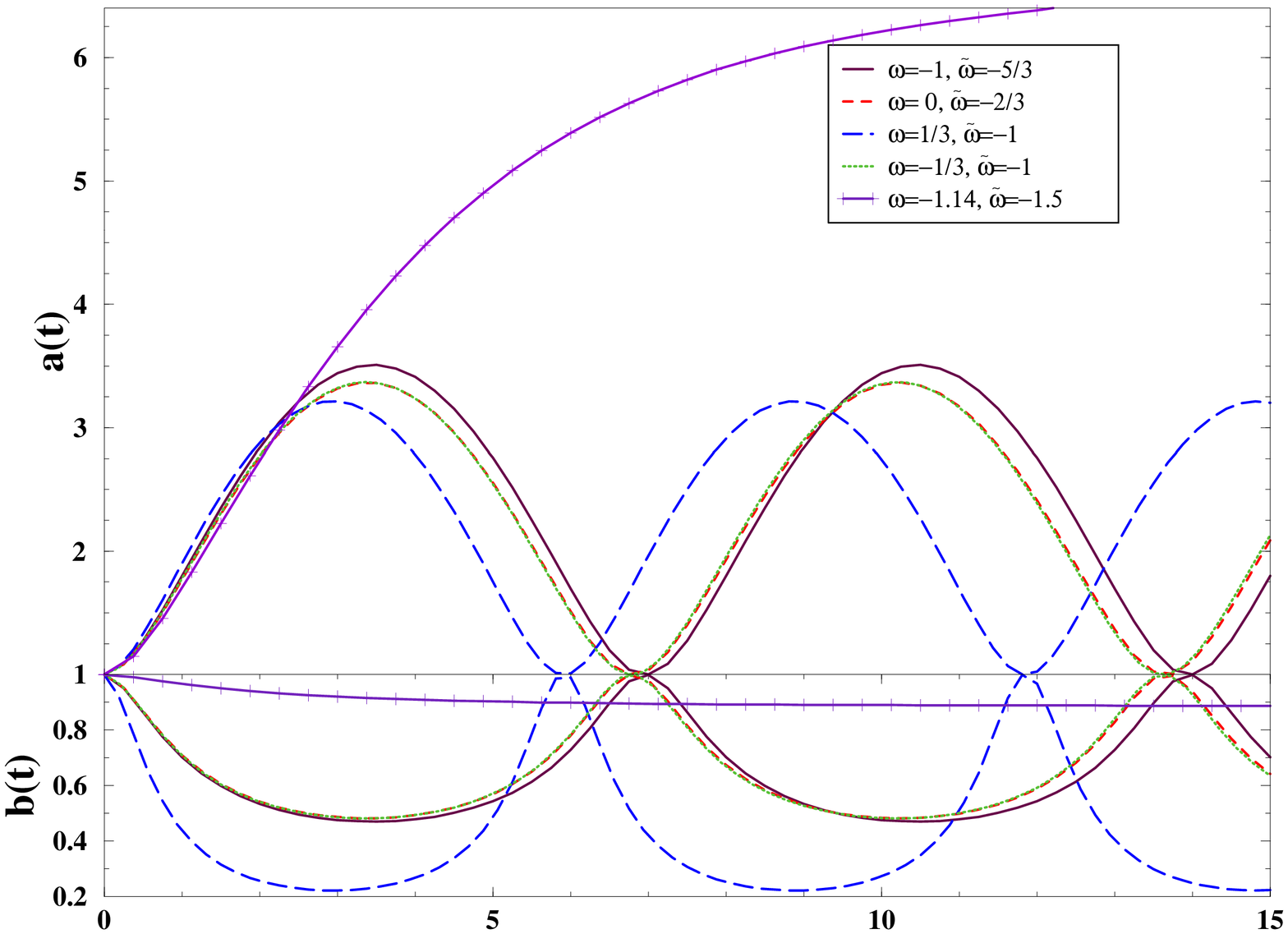}
\caption{Time evolution of the scale factors $a(t)$ and $b(t)$ in
an oscillating universe for $\Lambda_{6}<0$ and $k=-1$.}
\end{figure}
%%%%%%%%%%%%%%%%%%%%%%%%%%%%%%%%%%%%%%%%%%%%%%

Again these solutions are obtained only for negative values of
$\hat{w}$ according to Table 1 for various values of $w$. For $w<
-1$ there is a small range of values near the critical value of
$\hat{w} $, where $a(t)$ escapes from the oscillating behaviour
and grows very fast, forcing $b(t)$ to decay and freeze out at
certain small value analogously to De Sitter cases.

\subsection{Minkowski Bulk}

When $\Lambda_{6}$=0, we do not have the very strong effect of the
bulk cosmological constant and the time evolution of the scale
factors is smoother. If we had introduced a brane tension on the
4-brane, then a fine tuning similar to the Randall-Sundrum case in
five-dimensions, would have resulted in the same behaviour. We
show a typical time evolution of the two scale factors in Fig. 4,
for $k=0,-1$ while there is no acceptable solution for $k=1$.

%%%%%%%%%%%%%%%%%%%%%%%%%%%%%%%%%%%%%%%%%%%%%
\begin{figure}[h]
\centering
\includegraphics[scale=0.6,angle=-90]{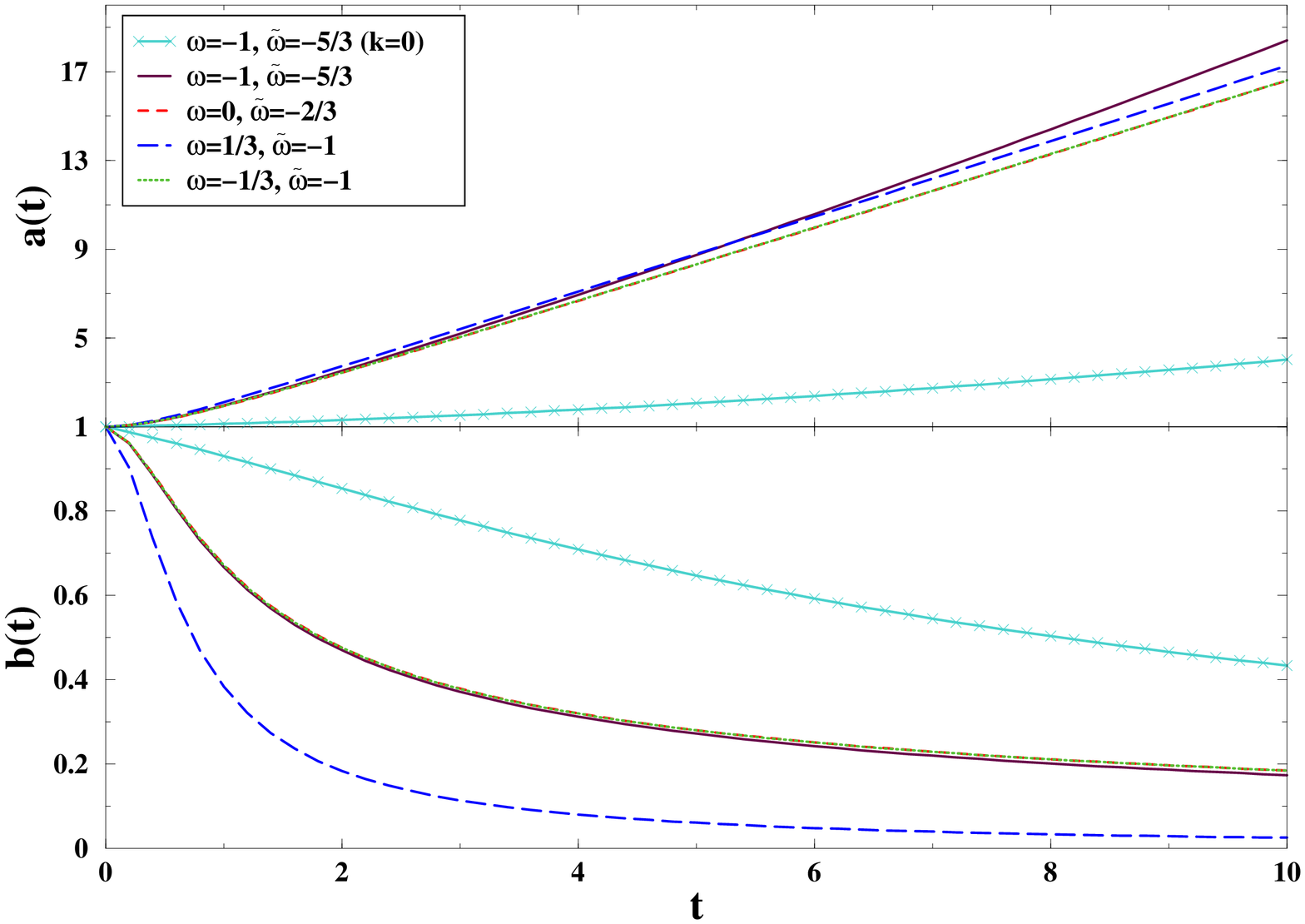}
\caption{Time evolution of the scale factors $a(t)$ and $b(t)$ for
$\Lambda_{6}=0$ and $k=0,-1$.}
\end{figure}
%%%%%%%%%%%%%%%%%%%%%%%%%%%%%%%%%%%%%%%%%%%%%%

The evolution of the scale factor $a(t)$ is nearly linear for
$k=-1$, while for $k=0$ there is an interesting solution with dark
energy in the physical universe ($w=-1$) and with "phantom" energy
in the extra dimension ($\hat{w}=-5/3$). The $b(t)$ scale factor
decays at a smaller rate compared with the case of a non-zero bulk
cosmological constant, but soon it gets a small value where it
freezes for the whole cosmological evolution. As it happens in all
the other cases, $\hat{w}$ is negative in the range of values
given in Table 1 for all acceptable solutions, indicating the need
of dark energy to suppress the extra fourth dimension compared to
the three other dimensions.

\section{Conclusions}

We presented a (4+1)-braneworld cosmological model in a
six-dimensional bulk. If $a(t)=b(t)$, with $a(t)$ the usual scale
factor of the three physical dimensions, and $b(t)$ the scale
factor of the extra fourth dimension, we found the generalized
Friedmann equation in six-dimensions of the Randall-Sundrum model
describing the cosmological evolution of a four-dimensional
brane-universe. We then showed that there is an equivalent
description of a 4-brane moving in a static six-dimensional bulk
and the cosmological evolution on the brane is described by the
same generalized Friedmann equation.

If $a(t) \neq b(t)$ the four-dimensional universe evolves with two
scale factors. However, for an observer in the moving brane, $a$
and $b$ are static depending only on the coordinate on which the
4-brane is moving. Then, demanding to have an effective
Friedmann-like equation on the brane, we showed that the motion of
the 4-brane in the static bulk is constrained by Darmois-Israel
boundary conditions, resulting in a relation
 connecting $a$ and $b$ acting as a constraint of the brane motion.

 The way $a$ and $b$ are related depends on
the energy-matter content of the 4-brane, because their relation
was derived from the consistency of the Darmois-Israel boundary
conditions. We then explored what are the consequences of the
presence of this constraint of motion for the cosmological
evolution of the brane-universe. We assumed that the universe
started higher-dimensional at the Planck scale with all the
dimensions at the Planck length, and subsequently an anisotropy
was developed between the three physical dimensions and the
extra-dimension. We then followed the evolution by making a
``phenomenological" analysis of how the two scale factors evolved
under various physical assumptions. In all considered cases, (A)dS
and Minkowski six-dimensional bulk, open, closed and flat
brane-universe and matter, radiation, cosmological constant and
dark energy dominated three-dimensional physical universe, we
found that dark energy is needed for the dynamical suppression and
subsequent freezing out of the extra fourth dimension.

It is interesting to further explore the relation we found between
the cosmological evolution of a higher-dimensional brane-universe
with the static properties of the bulk. The cosmological evolution
on a higher-dimensional brane-universe is related to a topology
change of the bulk during the evolution, and this relation might
lead to a better understanding of the
Gregory-Laflamme~\cite{Gregory:1993vy} instabilities of
higher-dimensional objects.

\section*{Acknowledgments}

We would like to thank E. Abdalla, A. Kehagias and R. Maartens for
useful discussions and BCM would like to thank NTUA for its
hospitality. This work was partially supported by the Greek
Education Ministry research program ``Pythagoras" and by Funda\c
c\~ao de Amparo \`a Pesquisa do Estado de S\~ao Paulo (FAPESP),
Brazil.

\end{document}